\documentclass[aps,pra,twocolumn,superscriptaddress]{revtex4-1}   	

\usepackage{graphicx}
\graphicspath{C:/Users/Soumya/Google Drive/Killian Lab/BEC_rydberg_lifetimes/BEC_rydberg_lifetime/paper/2020-09-01_paper-skk_praformat/figures/}
\usepackage{amssymb}

\begin{document}

\title{Loss rates for high-$n$, $49\lesssim n \lesssim150$, 5sns($^{3}$S$_{1}$) Rydberg atoms excited in an $^{84}$Sr Bose-Einstein condensate}

\author{S. K. Kanungo}

\author{J. D. Whalen}

\author{Y. Lu}

\author{\\T. C. Killian}

\author{F. B. Dunning}

\affiliation{Department of Physics and Astronomy, Rice University, Houston, TX  77005-1892, USA}

\author{S. Yoshida}

\author{J. Burgd\"{o}rfer}

\affiliation{Institute for Theoretical Physics, Vienna University of Technology Vienna, Austria, EU}

\begin{abstract}
Measurements of the loss rates for strontium n$^{3}$S$_{1}$ Rydberg atoms excited in a dense BEC are presented for values of principal quantum number $n$ in the range $49\lesssim  n \lesssim 150$ and local atom densities of $\sim1$ to $3\times10^{14}$~cm$^{-3}$.  Two main processes contribute to loss, associative ionization and state-changing.  The relative importance of these two loss channels is investigated and their $n$- and density-dependences are discussed using a model in which Rydberg atom loss is presumed to involve a close collision between the Rydberg core ion and ground-state atoms.  The present measurements are compared to earlier results obtained using rubidium Rydberg atoms.  For both species the observed loss rates are sizable, $\sim10^{5}-10^{6}$~s$^{-1}$, and limit the time scales over which measurements involving Rydberg atoms immersed in quantum degenerate gases can be conducted.
\end{abstract}

\maketitle

\section{Introduction}
\label{S:into}
Ultracold Rydberg molecules, which comprise a Rydberg atom in whose electron cloud are embedded one, or more, ground state atoms weakly bound through scattering of the Rydberg electron, are the subject of increasing interest due to their novel physical and chemical properties \cite{eil19,gds00}.    Initial experiments focused on the creation of dimer molecules comprising just one ground-state atom bound to a spherically-symmetric Rb(nS) Rydberg atom~\cite{bbn09}.  Measurements have since been extended to include a variety of different atomic species and atomic states~\cite{kgb14,trb12,dad15,amr14,bcb13}, and have demonstrated the formation of molecules containing multiple bound ground-state atoms, i.e., trimers, tetramers, ...\cite{lpr11,bbn11,gkb14}.  More recently, studies of ``molecule'' formation and loss in Bose-Einstein condensates (BECs) have been reported where the very-high ground-state atom densities (approaching 10$^{15}$ atoms cm$^{-3}$) allow the production of Rydberg atoms whose electron orbits can enclose hundreds or even thousands of ground-state atoms~\cite{sle16,ked18,wcd17,wcd17a}.  This work has demonstrated that such species provide an opportunity to probe the properties of cold dense gases~\cite{wkd19,dkw20}, to image the electron wavefunction~\cite{lse16,kbr15}, and to examine many-body phenomena such as the creation of Rydberg polarons in quantum degenerate gases~\cite{csw18}.
Such investigations, however, require that the lifetime of the Rydberg molecules be sufficiently long as to allow, for example, interactions to produce measurable
effects.  Initial studies of the lifetimes of Rydberg atoms excited in a BEC were
undertaken using Rb(nS) states created in a rubidium BEC and yielded lifetimes
of $\sim1-10\mu$s~\cite{slc16,sle16,ked18}.  The reactions responsible for
atomic loss, which involve a collision between the Rydberg core ion and nearest-neighbor ground-state atom, were investigated and identified.  More recently the lifetimes of
strontium 5sns($^{3}$S$_{1}$) Rydberg atoms with $n=49-72$ produced in an
$^{84}$Sr BEC have been explored~\cite{wcd17,wcd17a}.  Again lifetimes of
$\sim1-10 \mu$s were observed.

The collision of a Rydberg atom with a
neutral particle in a thermal gas has been intensively studied~\cite{bl95}.  The dynamics can be analyzed in terms of three different
processes: scattering of the Rydberg electron from the neutral particle, the acceleration of the
Rydberg core ion as a result of a collision with the neutral particle,
and a three-body collision involving the Rydberg electron
and the quasi-molecular ion formed by the Rydberg core and the neutral atom.
For ultracold gases
the energy exchange resulting from scattering of the Rydberg electron
by a ground state atom is  very small, and is insufficient to induce a transition
from the $5sns$($^{3}$S$_{1}$) state to another nearby Rydberg level.
Similarly,  the acceleration of the core ion during a collision
with a neutral atom will transfer momentum mostly to the center
of mass motion of the Rydberg atom and the transfer to
the electronic excitation (i.e., relative motion) is small.
When a ground state atom approaches the Rydberg core ion, however, an electric dipole moment
is induced in the atom and the corresponding interaction potential is
\begin{equation}
V_{pol} (R) = - \frac{\alpha}{2 R^4}
\label{eq:polarization}
\end{equation}
where $\alpha$ is the polarizability of the ground-state atom
($\alpha \simeq 186$~a.u. for strontium) and $R$ is the internuclear distance.
At low temperatures, the resulting attractive force accelerates the ion-atom pair towards each other resulting in a close collision.  (For the present range of $n$ and BEC densities, the separation between the Rydberg core ion and nearest-neighbor ground-state atom is much less than the size of the electron orbit and screening of the core ion charge is small.)
Loss of energy by the Rydberg
electron during such a close collision can lead to state-changing reactions
\begin{equation}
\label{eq:L-changing}
\text{Sr}(n^{3}S_{1})+\text{Sr}\rightarrow \text{Sr}(n^{'}L^{'})+\text{Sr}
\end{equation}
in which the Rydberg electron transitions to a neighboring lower-lying Rydberg level.
The energy released, $\sim 1/n^{3}$ a.u., is communicated to the
quasi-molecular ion-neutral atom pair.
Following such an interaction
the electron remains bound to the core ion producing
a ``fast'' Rydberg atom that quickly leaves the BEC.
The
Rydberg electron may also gain energy during a close collision through de-excitation of the transient Sr$_2^+$ ion,  leading to associative ionization
\begin{equation}
\label{eq:associative ionization}
\text{Sr}(n^{3}S_{1})+\text{Sr}\rightarrow \text{Sr}_{2}^{+} + e
\end{equation}

In the present work we extend earlier measurements in strontium to much larger values
of $n$, $n\sim150$, to explore the $n$ and local-density dependences of the
Rydberg loss rates. In addition, the present measurements are compared to
those reported previously using Rb($nS$) states.  Due to the absence
of a low-energy $p$-wave electron scattering resonance, the collision dynamics
for strontium are easier to analyze and can be described to a good approximation by a single Born-Oppenheimer potential surface.
Because  reactions~(\ref{eq:L-changing}) and~(\ref{eq:associative ionization})           can only occur when the separation between the core ion and
nearest-neighbor ground-state atom becomes small compared to the size of the Rydberg atom, their reaction rates can be analyzed using the time required for the Rydberg core ion to collide with the nearest-neighbor ground-state atom as starting point. This collision time
can be estimated using the Langevin model for ion-atom collisions.
The present observations show that the loss rates are approximately proportional
to density. Additionally,  the rates associated with associative ionization
(reaction~\ref{eq:associative ionization}) scale simply as $\sim1/n^{3}$
analogous to the rate predicted for a thermal gas~\cite{bl95}.
The loss rates associated with state-changing (reaction~\ref{eq:L-changing})
display a more complex $n$-dependence. For
values of $n\gtrsim110$, the present measured loss rates agree well with those
reported previously for rubidium.  At lower values of $n$, however, the loss
rates seen for strontium are considerably smaller than those for rubidium
highlighting the role played by the low-energy $p$-wave electron scattering
resonance in rubidium.  Nonetheless, for both species the observed loss rates
are sizable and limit the time scales over which measurements involving
Rydberg atoms immersed in quantum degenerate gases can be conducted.

\section{Experimental Method}
 \label{S:exp method}
 The present experimental approach has been described in detail previously~\cite{ssk14,mmy09}.  Briefly, $^{84}$Sr atoms are first cooled in a ``blue'' magneto-optical trap (MOT) using the 5s$^{2}$ $^{1}$S$_{0} \rightarrow$ 5s5p$^{1}$P$_{1}$ transition at 461~nm following which they are further cooled in a ``red'' MOT operating on the narrow 5s$^{2}$ $^{1}$S${_0} \rightarrow$5s5p $^{3}$P$_{1}$ intercombination line at 689~nm.  The atoms are then loaded into an optical dipole trap (ODT) formed by two crossed 1.06~$\mu$m laser beams where they are subject to evaporative cooling to create a BEC.   The peak trap densities are determined from measurements of the total atom number and trap oscillation frequencies.   Typically $\sim2-8\times 10^{5}$ atoms are trapped with peak densities of up to $\sim3\times10^{14}$~cm$^{-3}$.  The temperature of the atoms is estimated to be $\sim120$~nK and the condensate fraction is $\gtrsim95\%$.  Stray fields in the ODT are minimized by the application of small offset potentials to a series of electrodes that surround the ODT which allows excitation of well-defined Rydberg states with values of $n\lesssim150$.

 Rydberg atoms are created by two-photon excitation via the 5s5p $^{3}$P$_{1}$ intermediate state.  The 689-nm laser for the first step is tuned 80~MHz to the blue of the intermediate state to reduce single-photon scattering.  The 319-nm laser required for the second step is tuned to the Rydberg state of interest.  Both excitation lasers are stabilized to the same high-finesse optical cavity.  The detuning, $\Delta E$, of the laser from the unperturbed atomic state determines the local density, $\rho$, in which a Rydberg atom is excited.  For large ground-state atom densities, $\Delta E$ and $\rho$ are approximately related by the mean field expression
 \begin{equation}
 \label{eq:mean field}
 \Delta E =\frac{2\pi \hbar^{2}}{m_{e}}a_{s_{eff}}\rho
 \end{equation}
 where $m_{e}$ is the electron mass and $a_{s_{eff}}\sim-11a_{0}$ is the effective s-wave scattering length which reflects the average of the interactions over the Rydberg wave function.  The value of $a_{s_{eff}}$ is $n$ dependent but varies by $\lesssim10\%$ for the present range of $n$~\cite{swd18}.  The excitation lasers cross at right angles and have circular and linear polarizations, respectively, leading to creation of 5sns $^{3}$S$_{1}$ states with magnetic quantum number $M_{J}=1$.  The excitation lasers are chopped to form a periodic train of optical pulses, each of $\sim 2 \mu$s duration, with a pulse repetition frequency of $\sim4$~kHz.  The ODT is turned off during excitation to eliminate ac Stark shifts.  The number of Rydberg atoms created during a single pulse is limited to $\lesssim0.2$ to eliminate any possibility of effects due to blockade.

 Following excitation the number and excited-state distribution of the surviving Rydberg atoms is measured as a function of the delay time, $t_{D}$, from the end of the laser pulse using selective field ionization (SFI)~\cite{wcd17a,sdu83,gal94} for which purpose the atoms are subject to a time-dependent electric field of the form $E(t)=E_{p}(1-e^{-t/\tau})$ with a time constant $\tau\sim6\mu$s.  Ionization, however, typically begins $\sim1\mu$s from the start of the field ramp.  Electrons produced by SFI are directed to a microchannel plate (MCP) detector whose output pulses are fed to a multichannel scaler (MCS).  The electric field at which ionization occurs is determined from the electron arrival time at the MCP and the time dependence of the applied electric field.  Data are accumulated following many laser pulses to build up good statistics.  Approximately $4\times10^{3}$ experimental cycles can be performed using a single BEC sample.

\begin{figure}
  \includegraphics[scale = 0.9]{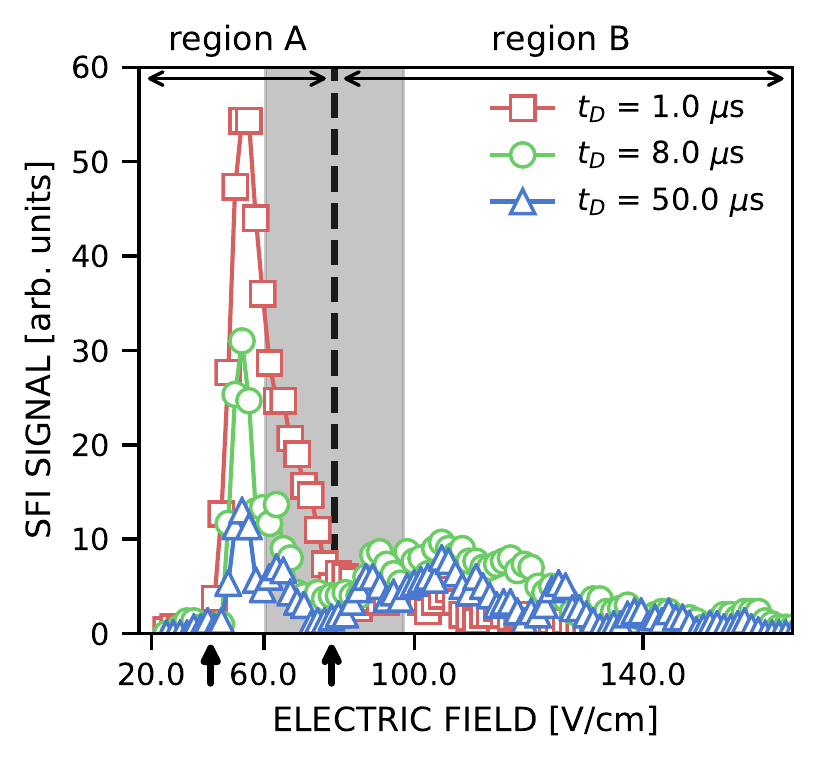}
 \caption{\label{fig:SFI spectra}
   SFI spectra recorded for $n=55$ Rydberg atoms excited in a BEC with local atom density $\rho =8\times10^{13}$~cm$^{-3}$, and the delay times $t_{D}$ indicated.  The vertical arrows indicate the threshold fields for adiabatic ionization (1/16$(n-\delta)^{4}$ ~a.u.) and diabatic ionization (1/9$(n-\delta)^{4}$~a.u.).  The dashed line separates the regions of interest A and B discussed in the text. The shaded region indicates the range of positions of this dividing line that result in a less than $\pm5\%$ change in the extracted loss rates (see text).}
\end{figure}

 Figure \ref{fig:SFI spectra} shows SFI spectra recorded at several values of delay time using $n=55$
Rydberg atoms and a local atom density $\rho \sim 8\times10^{13}$~cm$^{-3}$.
The field at which a particular Rydberg state ionizes depends on $n, \vert
M_{L}\vert$,  and the slew rate of the applied field.  The relatively narrow
feature seen at low fields at the shortest delay time results primarily from
ionization of parent $M_{J} =1$ ($M_{L}=0$) Rydberg atoms along a mostly adiabatic
path involving a multitude of avoided crossings~\cite{wcd17}
leading to
ionization across the Stark saddle with a threshold
given by
$\sim1/16(n-\delta)^{4}$~a.u., where $\delta=3.371$ is the quantum defect, and
corresponds to a field of $\sim45$~V cm$^{-1}$.  With increasing time delay $t_D$,
the total SFI signal decreases and the form of the SFI spectrum changes
dramatically.  The ionization signal at the lower fields decreases
substantially, this being accompanied by the growth of a signal that
encompasses a range of higher field strengths that correspond to ionization of
higher-$\vert M_{L}\vert$ states along principally diabatic paths~\cite{wcd17}
that resemble the ionization paths of the most red shifted hydrogen Stark states and for which the threshold is given by $\sim1/9(n-\delta)^{4}$, i.e.,
$\sim80$~V~cm$^{-1}$.  In practice, for each value of $n$, the size of the peak
applied field $E_{p}$ was varied $(\sim n^{-4})$
such that the adiabatic and diabatic ionization signals arrived
at the same relative times during the ramp and were
similarly well resolved.  Earlier work suggests that for $n \simeq 60$
and electric field slew rate of 43~V/cm/$\mu$s, states with $|
M_{L} | \leq 3$ tend to ionize along principally adiabatic paths, whereas
states with larger values of $| M_{L} |$ tend to ionize along
predominantly diabatic paths~\cite{wcd17a}.  As $n$ increases, however, the low $\vert M_{L}\vert$ states follow increasingly non-adiabatic paths to ionization.  Nevertheless, the SFI spectra recorded at the
other values of $n$ studied in this work exhibit similar characteristics to
those seen in Fig.~\ref{fig:SFI spectra}.  In general, as the delay time
increases the total surviving Rydberg population decreases and there is an
evolution towards states of higher $\vert M_{L}\vert$, i.e., higher $L$.

\section{Rate equation for the ionization dynamics}
\begin{figure*}
   \includegraphics[width = 0.95\textwidth]{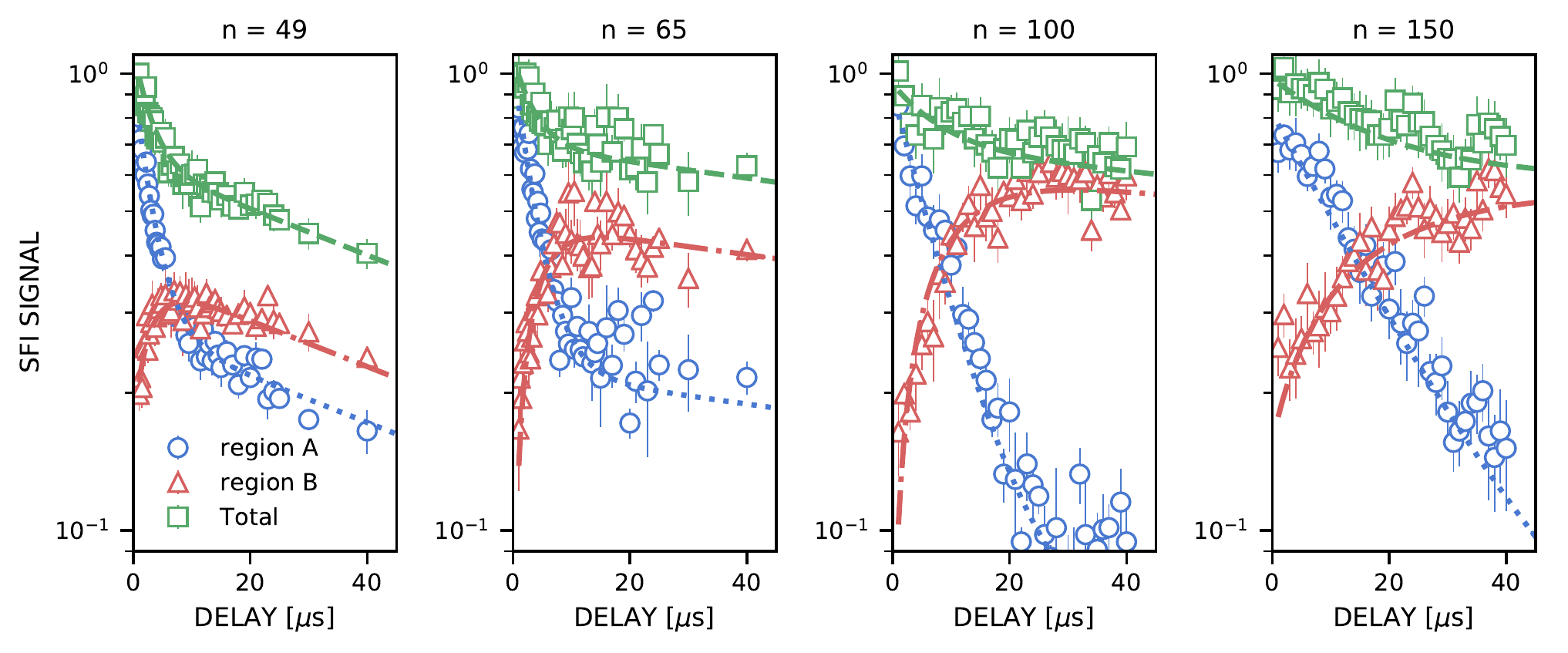}
 \caption{\label{fig:time evolution}
 Measured time evolution of the total SFI signal as a function of delay time
 $t_{D}$ for the values of $n$ indicated.  The figure also includes the
 separate evolutions of the SFI signals in regions A and B.  The lines
 show fits to the data obtained using Eqs.~(\ref{eq:rate equation})-(\ref{eq:B
   population}).  The local atom density $\rho$ is
 $8\times10^{13}$~cm$^{-3}$.}
\end{figure*}

In a BEC two processes primarily contribute
to the loss of parent Rydberg atoms: associative ionization and
Rydberg state-changing transitions (henceforth simply referred to as $L$-changing reactions) both of which require energy exchange in an interaction
between the (transient) Sr$_2^+$ ion formed by the collision of the core ion and nearest-neighbor ground-state atom and the Rydberg
electron.
To extract information on the dynamics of associative ionization and $L$-changing
from the measured SFI signals, the time evolution of the Rydberg atom population
is approximated using the rate equations
\begin{eqnarray}
&& \frac{dN_{p}}{dt} = -(\Gamma_{AI}+ \Gamma_{L} + \Gamma_{R}) N_{p}
\nonumber \\
&& \frac{dN_{L}}{dt}= \Gamma_{L} N_{p}-\Gamma_{R}N_{L}
\label{eq:rate equation}
\end{eqnarray}
where $N_{p}$ is the number of parent Rydberg atoms and
 $N_L$ is the number of $L$-changed Rydberg atoms.
$\Gamma_{AI}$ is the rate for associative ionization,
$\Gamma_{L}$ is the rate for $L$-changing collisions,
and $\Gamma_{R}$ is the average radiative decay rate
which includes the effects of interactions with background blackbody
radiation. For simplicity, it is assumed that the radiative decay rates
for Rydberg atoms are  independent of $L$.
Measurements of the decay rates for Rydberg atoms created in (low-density)
thermal samples and (as will be demonstrated) for $L$-changed atoms created in
a BEC yield similar rates $\Gamma_{R}\sim 5\times 10^{3}-10^{4}$~ s$^{-1}$ which
are much smaller than the decay rates associated with associative
ionization or $L$-changing, i.e., $\Gamma_R \ll \Gamma_{AI}, \Gamma_L$. Analytic solution of the rate equations
[Eq.~(\ref{eq:rate equation})]
with the boundary condition $N_{p}(t=0) = N_0$ and $N_{L}(t=0)=0$ yields
\begin{eqnarray}
 \label{eq:rate equation solution}
&& N_{p}=N_{0} ~e^{-(\Gamma_{AI}+\Gamma_{L}+\Gamma_{R})t}
\nonumber \\
&& N_{L}=N_{0}~\frac{\Gamma_{L}}{\Gamma_{AI}+\Gamma_{L}}~e^{-\Gamma_{R}t}
  [1-e^{-(\Gamma_{AI}+\Gamma_{L})t}] \, .
\end{eqnarray}
Although the $L$-changed atoms quickly exit the BEC, they remain sufficiently close to the BEC during an experimental cycle that they are still detected, whereupon $L$-changing does not effectively contribute to the loss of total SFI signal.  The values of $\Gamma_{AI}$ and $\Gamma_{L}$ can be obtained
by fitting the measured total SFI signals, shown in Fig.~\ref{fig:time evolution},
 to the analytical solution $N_{p}(t) + N_{L}(t)$.

In the present work we adopt a different approach
with the goal to disentangle in more detail the two rates, $\Gamma_{AI}$ and
$\Gamma_L$.
As illustrated in Fig.~\ref{fig:SFI spectra}, the
SFI spectra are divided into two regions of interest:
the SFI signal appearing at low electric fields, i.e., in region A, and at higher electric fields, i.e., in region B. For each value of $n$, the boundary between these regions
is taken to be the threshold for diabatic ionization,
$1/(9(n-\delta)^4)$. Ionization in region A
results primarily from ionization of the parent Rydberg atoms plus a
contribution from the low-$\vert M_{L}\vert$ $L$-changed collision products.
Region B contains the high-$\vert M_{L}\vert$ $L$-changed collision
products and a small fraction of the parent atoms.
The populations $N_{A}$ and $N_{B}$ in regions A and B are
individually fit to
\begin{equation}
 \label{eq:A population}
 N_{A} = (1-\epsilon_{p})N_{p} + \epsilon_{L}N_{L}
 \end{equation}
 \begin{equation}
 \label{eq:B population}
 N_{B} = (1-\epsilon_{L})N_{L} + \epsilon_{p}N_{p}
\end{equation}
where $\epsilon_{p}$ and $\epsilon_{L}$ are the fractions of
the parent and $L$-changed populations that ionize in regions A and B,
respectively. $\epsilon_{P}, \epsilon_{L}$, $N_{p}$ and $N_{L}$, together with the decay rates
$\Gamma_{AI}$, $\Gamma_{L}$, and $\Gamma_{R}$, are obtained through
fits to the experimental data. Tests revealed that the values of $\Gamma_{AI}$ and $\Gamma_{L}$ obtained by such fitting
are not sensitive to the exact positioning of the boundary between regions A and
B. For example, placing the
boundary at different positions within the region shaded in gray in
Fig.\ref{fig:SFI spectra} resulted in less than a $\pm5\%$ change in the
fitted values of $\Gamma_{AI}$ and $\Gamma_{L}$.  The values of $\Gamma_{AI}$
and $\Gamma_{L}$ obtained in this manner matched well with those obtained by
fitting the total SFI signals.
Figure~\ref{fig:time evolution} displays the measured evolution of
$N_A$ and $N_B$ as a function of the time delay $t_D$.
For $n$=49, at early times the signal in region A decreases near exponentially due to rapid associative ionization.  The rate of this decrease moderates at later times due to the increasing contribution from long-lived low-$\vert M_{L}\vert$ products of $L$-changing reactions.  In contrast, the signal in region B initially grows due to the creation of high-$\vert M_{L}\vert$ $L$-changed collision products.  At late times, both signals (which are each associated with $L$-changed products) decay at similar slow rates.  This rate (which is essentially the same for all values of $n$) corresponds to a radiative decay rate $\Gamma_{R}\sim10^{4}$~s$^{-1}$.

\section{Results and Discussion}
 \label{S:Results}
\begin{figure}
  \includegraphics[scale = 0.95]{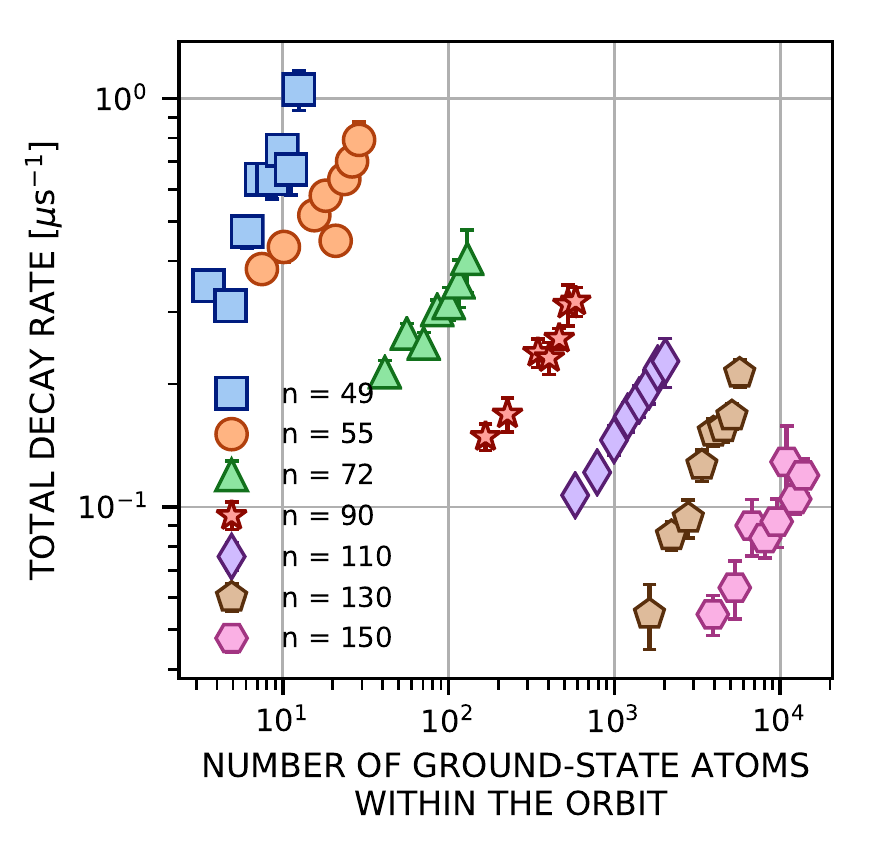}
\caption{\label{fig:total decay}
Measured total loss rates, $\Gamma_{total}=\Gamma_{AI}+\Gamma_{L}+\Gamma_{R}$, for the values of $n$ indicated as a function of the number of ground-state atoms within the Rydberg electron cloud.}
\end{figure}

\begin{figure*}
   \includegraphics[width = 0.95\textwidth]{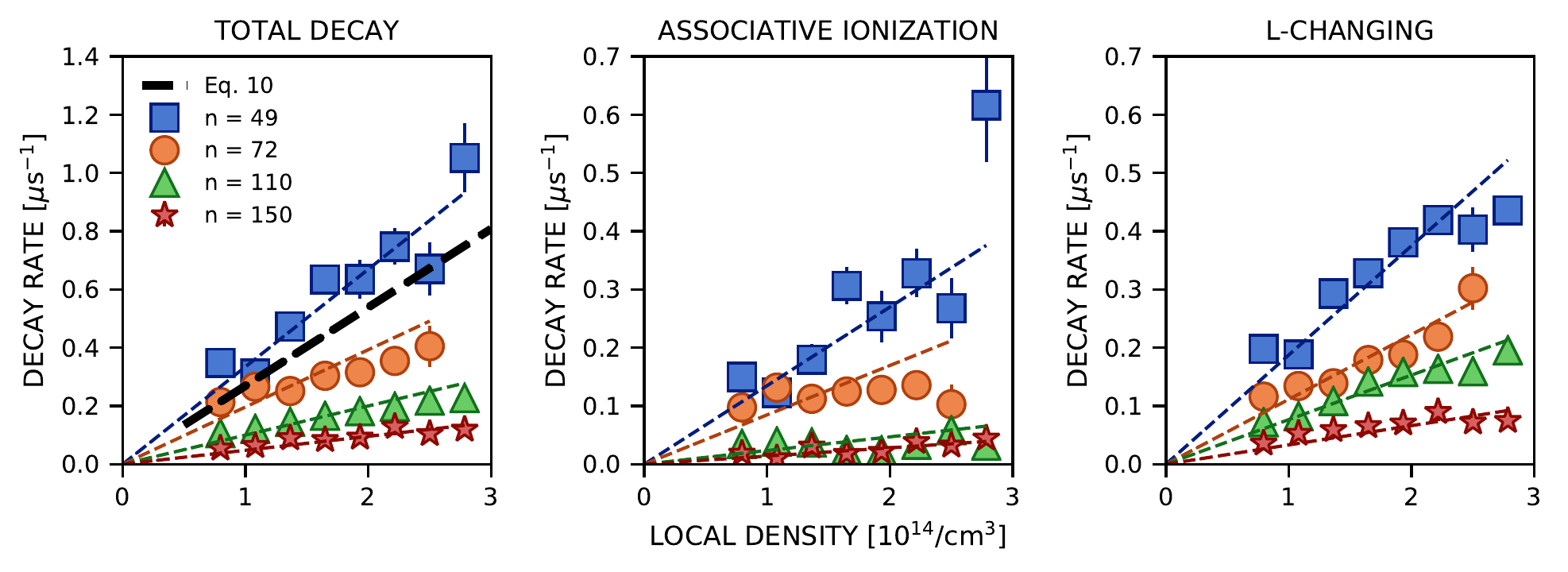}
 \caption{\label{fig:measured total loss}
 Measured total loss rates $\Gamma_{total}$, loss rates for associative ionization ($\Gamma_{AI}$), and rates for $L$-changing ($\Gamma_{L}$) as a function of local atom density $\rho$ for the representative values of $n$ indicated.  The lines show linear fits to the data.  Loss rates calculated using Eq.~\ref{eq:collision rate} (---) are also included .}
\end{figure*}

Figure~\ref{fig:total decay}
shows the measured total decay rates ($\Gamma_{total}=\Gamma_{L}+\Gamma_{AI}+\Gamma_{R}$) for
several values of $n$ expressed as a function of the total number of ground
state atoms contained within the Rydberg electron cloud, given by $4\pi
r^{3}\rho/3$, where $\rho$ is the local density and $r\sim2n^{2} a_{0}$.  Two
features of the data are immediately apparent.  Firstly, for a given number of
atoms within the electron orbit the total decay rate decreases with increasing
$n$. For example, with about 1500 ground state atoms residing within the Rydberg
electron cloud, the total decay rate for $n=110$ is a factor of 3 to 4 times larger than
for $n=130$.
Secondly, for a given value of $n$, the loss rate increases steadily as
the number of ground-state atoms in the electron cloud increases.  This is
further illustrated in Fig.~\ref{fig:measured total loss}
which displays measured total loss rates as a function of laser detuning, which is directly proportional to $\rho$ (Eq.~\ref{eq:mean field}), for
several values of $n$.  As seen in Fig.~\ref{fig:measured total loss}, the
measured loss rates for each value of $n$ scale linearly with $\rho$.

In the following we analyze the rates for
associative ionization and $L$-changing separately.
When the initial separation, $R_i$, between the Rydberg core ion
and nearest-neighbor ground state atom is substantially smaller than the size of
the electron orbit, their interaction is governed by
the polarization potential $V_{pol}(R)$ (Eq.~\ref{eq:polarization}).
For a ground state atom with energy near the dissociation threshold,
the collision time can be
estimated~\cite{wcd17,wcd17a} as
\begin{equation}
t_{col} \simeq \frac{1}{3} \sqrt{\frac{\mu}{\alpha}} R_i^3
\end{equation}
where $\mu$ is the reduced mass of the collision pair.
If only the pair interaction between the core ion and neighboring ground-state atom is considered, $\mu=M/2$ where $M$ is the $^{84}$Sr mass.  However, when many ground-state atoms are bound to the core ion in the molecular potential the effective reduced mass might be somewhat larger.  An average collision rate for a binary ion core-ground state atom collision can be estimated by considering a uniform distribution of ground state atoms as
\begin{equation}
\langle \Delta t_{col} \rangle^{-1} \simeq
4 \pi \rho \sqrt{\frac{\alpha}{\mu}}.
\label{eq:collision rate}
\end{equation}
$\langle \Delta t_{col} \rangle$ is equivalent to the
average time required for a collision with the nearest neighbor
and, assuming $\mu=M/2$, is estimated to be $\sim 3.2~\mu$s for $\rho = 8 \times 10^{13}$~cm$^{-3}$. This corresponds to a collision rate of
$\langle \Delta t_{col} \rangle^{-1} \simeq 3.1\times10^{5}$~s$^{-1}$.
This rate should provide an upper bound to the reaction rate governing
the decay of Rydberg atoms. Indeed, this rate is close to the observed total decay rate
$\Gamma_{total}$ for $n=49$ at this density (Fig.~\ref{fig:measured total loss}).
For the higher values of $n$, the measured reaction rates are, however, much smaller than this binary collision rate.
In addition, as seen in Fig.~\ref{fig:time evolution}, reactions occur on time scales much longer than $\langle\Delta t_{coll}\rangle$.  Taken together, these observations suggest that multiple core ion-ground state atom collisions may be required before reaction occurs, the overall reaction rate being given by the product of the collision rate and the probability that each collision will result in a reaction.  (Multiple collisions are possible because, following their initial collision, ion-atom pairs that do not react remain bound and can undergo further periodic collisions.  In addition, at late times collisions involving next-to-nearest-neighbor and even more distant ground-state atoms begin to occur.).

Extracting the probability that a single ion-atom collision will lead to reaction from the measured loss rates is challenging because the collision rate for a bound ion-atom pair depends on how strongly they are bound in the potential $V_{pol}(R)$ (Eq.~\ref{eq:polarization}), which depends on the many-body environment of nearby ground state atoms. For example, the binding energy of a given pair may change because scattering of the core ion during an ion-atom collision can lead to the loss of bound ground-state atoms from the initial Rydberg ``molecule'' as a result of which the ion-atom pair will become more strongly bound and their collision frequency will increase.  The sum of the binding energies of all the ground-state atoms equals the laser detuning $\Delta E$,  which is proportional to $\rho$ (see Eq.~(\ref{eq:mean field})).  As an example, if it is assumed that the binding energy of an ion-atom pair is increased by this amount, estimates show that, for a detuning $\Delta E=-5.6$~MHz, i.e., a density $\rho=8\times10^{13}$~cm$^{-3}$, the collision frequency will approach $\sim10^{7}$~s$^{-1}$.  This value is much higher than the measured loss rates suggesting that the probability of reaction during an ion-atom collision can become rather small.

\begin{figure}
   \includegraphics[scale = 0.95]{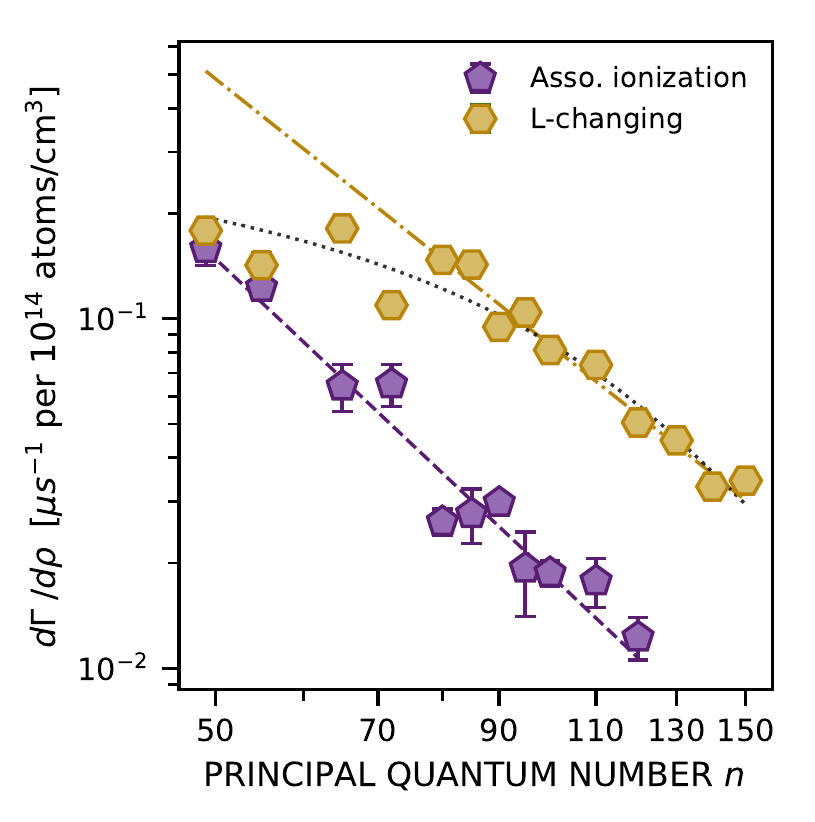}
 \caption{\label{fig:slopes}
 Measured $n$ dependence of the slopes $d\Gamma_{AI}/d\rho$ and $d\Gamma_{L}/d\rho$ (see text).  The dashed line is a fit to the results for associative ionization and scales as $1/n^{3.0}$.  The dash-dotted line is a fit to the $L$-changing data at high $n$ and scales as $1/n^{2.5}$.  The dotted line is drawn as a guide to the eye.}
 \end{figure}

Since the binary collision rate (Eq.~\ref{eq:collision rate}) is determined by the density, or equivalently, laser detuning, it is independent of $n$.  However, as is evident from Fig.~\ref{fig:measured total loss}, the total loss rates $\Gamma_{total}$ decrease steadily with increasing $n$ which must therefore be attributed to a decrease in the reaction probabilities during ion-atom collisions. The measured loss rates $\Gamma_{AI}$ and $\Gamma_{L}$ both increase linearly with the local density, $\rho$.  Therefore the reaction probabilities can be separated from the density-dependent collision rate (Eq.~\ref{eq:collision rate}) by analyzing the slopes, $d\Gamma /d\rho$, which are shown in Fig.~\ref{fig:slopes}.

We first consider associative ionization which requires
the autoionization of the Rydberg electron through an energy exchange with
the valence electron of the transient Sr$_2^+$ ion. Since the electron-electron
scattering probability is approximately proportional to the overlap
of the electron wavefunctions, it is reasonable to expect that the associated reaction probability will scale with the local Rydberg electron probability density near the valence electron which
 decreases rapidly with increasing $n$, scaling as $1/(n-\delta)^{3}$.
(The autoionization rate for thermal collisions between a Rydberg atom
and a ground state atom is known to scale as  $1/(n-\delta)^{3}$~\cite{bl95}).
As seen in Fig.~\ref{fig:slopes}, the measured values of $d\Gamma_{AI}/d\rho$ do indeed decrease rapidly
with increasing $n$ scaling as $\sim 1/n^{3.0}$, consistent with this picture.
(For values of $n\gtrsim110, \Gamma_{AI}$ and $d\Gamma_{AI}/d\rho$ become
too small to be reliably measured.)

In $L$-changing reactions the Rydberg electron exchanges energy
not with the valence electron of the transient Sr$_2^+$ ion
but with its vibrational motions.  Whereas energy exchange can be caused by acceleration of the ion core,
a momentum transfer $\Delta p$ to the ion core
will accelerate the Rydberg electron with respect to the ion core
by $(m_e/M) \Delta p$, resulting in only very small energy transfer to the Rydberg
electron. This suggests that the major contribution to energy
exchange most likely results from the interaction of the Rydberg electron with the electronic dipole moment
$\vec{D}(r)$ of the Sr$_2^+$ molecular ion.
The transition probability from a given $n$ level to an adjacent $n$ level
induced by the electron-dipole interaction is estimated to scale as
$\sim 1/n^2$~\cite{bl95} when contributions
from all initial $L$ and $M$ levels are considered.   At high $n$, $d\Gamma_{L}/d\rho$ is seen to scale as $\sim1/n^{2.5}$ (see Fig.~\ref{fig:slopes}) which is close to this predicted scaling, the discrepancy possibly resulting because here we only consider $L$-changing from an initial $S$-state.
At the lower values of $n$, however, the $n$-dependence of the $L$-changing rate
decreases dramatically and the rate becomes  comparable
to that for associative ionization.
A more detailed analysis of the
molecular potential at small distances will be required to quantify
the competition between associative ionization
and $L$-changing in this regime.

If $L$-changing results from dipole interactions it should, starting from parent $S$-states, preferentially populate low-$L$ states.
As is evident from Fig.~\ref{fig:time evolution},
for $n=49$ a sizable fraction, $\sim40\%$, of the product $L$-changed atoms
are in low-$\vert M_{L}\vert$ states, i.e., in region A.
This is considerably larger than expected for a statistical distribution of $n=49$ states for which fewer than 14\% of atoms are in states with $|M_L| \le 3$.
The data therefore point to the
preferential population of low-$\vert M_{L}\vert$, and hence low-$L$, states.
As $n$ increases the fraction of $L$-changed atoms that ionize in region A
decreases but again the data suggest that low-$L$ states are preferentially
populated.  For example, at $n$=100, $\sim17\%$ of the $L$-changed atoms
ionize in region A whereas this number should be $\sim9\%$ for a statistical mix
of states.

\begin{figure*}
  \includegraphics[scale = 0.95]{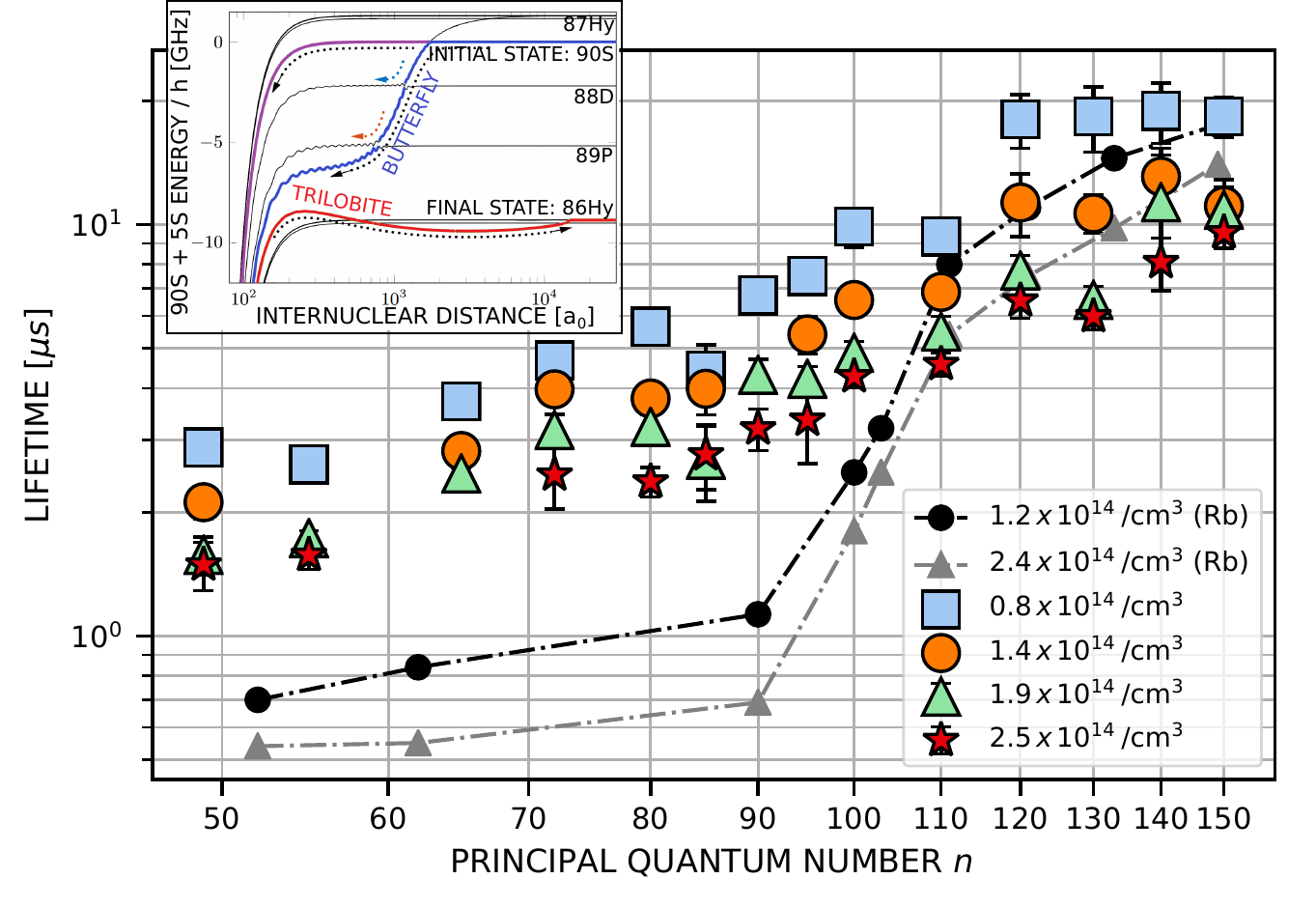}
\caption{\label{fig:comparison}
Comparison of the atomic lifetimes, $\tau=1/(\Gamma_{AI}+\Gamma_{L}+\Gamma_{R})$ measured in the present work with those reported previously for Rb(nS) molecules (see text) for varying local atomic densities as indicated.
The inset provides a schematic of the potential energy landscape for Rb(90S) and includes the ``butterfly" state associated with the p-wave shape resonance in electron scattering from rubidium.  Possible behaviors at the resulting avoided crossings are also indicated (adapted from ref~\cite{sle16}).}
\end{figure*}

It is interesting to compare the present results with those reported previously using Rb(nS) states~\cite{sle16}.
Figure~\ref{fig:comparison}
compares the present measured lifetimes, i.e., $1/(\Gamma_{AI}+\Gamma_{L}+\Gamma_{R})$, with those obtained in the earlier rubidium experiments.  The lifetime, $\tau$, of the rubidium states was obtained using the expression
\begin{equation}
\frac{1}{\tau}= \frac{1}{\tau_{Rb_{2}^{+}}} + \frac{1}{\tau_{L}}
\end{equation}
where $\tau_{Rb_{2}^{+}}$ is the $Rb_{2}^{+}$ formation time and $\tau_{L}$ the $L$-changing collision time (see ref~\cite{sle16}).
For large values of $n$, $n\gtrsim110$, the lifetimes measured in the present work are very similar to those seen in rubidium.  However, for values of $n\lesssim110$ the lifetimes become markedly different.  In particular, in the case of rubidium there is a pronounced drop in the lifetime between $n=110$ and $n=90$ followed by a slower decrease at lower $n$.   No similar sudden drop is seen in strontium.

The short lifetimes seen for $n\lesssim90$ in rubidium were attributed to the
$p$-wave shape resonance at $\sim0.03$~eV present in electron-rubidium
scattering which significantly affects the potential energy landscape leading
to the appearance of the so-called ``butterfly'' state.  As illustrated by the
Born-Oppenheimer potential surfaces shown in the inset in
Fig.~\ref{fig:comparison}, this results in a series of avoided crossings as
the separation between the ground-state atom and core ion decreases.
Calculations suggest that there is a sizable probability that upon approach a
parent nS state will transition to the ``butterfly'' state which accelerates
the core ion and neutral atom towards each other, enhancing the collision rates.
Since strontium does not possess such a $p$-wave
resonance, the observed longer lifetime of Sr Rydberg states
supports the claim that this scattering feature plays an important
role in determining the rubidium lifetime.  The similar lifetimes observed at high $n$ in rubidium and strontium suggest that the butterfly potential is not significantly influencing the loss in this regime in rubidium.  Furthermore, long lifetimes imply that multiple binary ion-atom collisions are required before reaction occurs.

\section{Conclusions}
\label{S:conclusions}
The present study demonstrates that strontium Rydberg atoms created in a dense
BEC undergo rapid associative ionization and $L$-changing processes that are
initiated by the attractive ion-induced dipole interaction between the
``bare" Rydberg core ion and nearest-neighbor ground-state atom which leads
to their collision.  Both mechanisms require the presence of the Rydberg
electron during the collision to facilitate energy exchange.
The destruction rate due to associative ionization decreases rapidly with increasing $n$, scaling  as $\sim1/n^{3.0}$ which matches the $n$-dependent decrease in the electron probability density near the ion-ground state atom collision pair.  The $L$-changing rate displays a more complex behavior.  The 1/$n^{2.5}$ scaling at high $n$ and the approximate $n$-independence at the lower quantum numbers suggest that the internuclear separation at which the $L$-changing events occur and the energy released during such processes both play non-trivial roles in determining the rate.

In future studies it will be of considerable interest to test the hypothesis that the loss rate is limited by the electron probability density near the core ion by undertaking measurements starting from high-$L$ Rydberg states for which this probability  is greatly reduced.  It will also be of interest to examine the properties of Rydberg atoms excited in dense spin-polarized $^{87}$Sr samples for which fermion statistics limit the likelihood of finding two ground-state atoms in close proximity which should again delay collisions and result in longer molecular lifetimes.

 \begin{acknowledgments}
This research was supported by the NSF under Grant No. 1904294, the AFOSR under Grant No. FA9550-17-1-0366, the Robert A. Welch Foundation under Grants
Nos. C-0734 and C-1884, the FWF(Austria) under Grants No. FWF-SFB041ViCom
and the FWF-doctoral college W1243.

 \end{acknowledgments}

\bibliography{references}


\end{document}